\documentclass[lnicst]{svmultln}
\usepackage{makeidx}  
\usepackage{comment}                    
\usepackage{algorithm}                  
\usepackage[noend]{algorithmic}         
\usepackage{graphicx}
\usepackage{subfigure}                  
\usepackage{amsmath}
\usepackage{epsfig}                     
\usepackage{graphicx}
\usepackage{epstopdf}
\begin{document}
\mainmatter              
\title{A Machine Learning Based   Forwarding Algorithm Over Cognitive Radios in Wireless Mesh Networks}
\titlerunning{Jianjun Yang et al.}  
%
\author{Jianjun Yang\inst{1},  Ju Shen\inst{2}
, Ping Guo \inst{3}, \\Bryson Payne\inst{1}  and Tongquan Wei\inst{4}}
\authorrunning{Jianjun Yang et al.}   
%
\tocauthor{Jianjun Yang, Ju Shen, Ping Guo}
\institute{ University of North Georgia, Gainesville, GA, USA,\\
\and
University of Dayton, Dayton, OH, USA
\and
University of Illinois at Springfield, Springfield, IL, USA
\and
East China Normal University, Shanghai, China}

\maketitle              

\begin{abstract}        
Wireless Mesh Networks improve their capacities by equipping mesh nodes with multi-radios tuned to non-overlapping
channels. Hence the data forwarding between two nodes has multiple selections of links and the bandwidth between the pair of nodes
varies dynamically. Under this condition, a mesh node  adopts machine learning mechanisms to choose the possible best
next hop which has maximum bandwidth when it intends to forward data.
In this paper, we present a
machine learning based   forwarding algorithm to let a forwarding node dynamically select the next hop with highest potential bandwidth
capacity to resume communication based on learning algorithm.
Key to this strategy is
that a node only maintains three past status, and then it is able to  learn and   predict the potential  bandwidth capacities of its links. Then, the
 node  selects the next hop with
potential maximal link bandwidth.
Moreover, a geometrical based algorithm is developed to let the source
node figure out the forwarding region in order to avoid  flooding.
Simulations demonstrate that our
approach significantly speeds up the transmission and outperforms other peer algorithms.

\end{abstract}
\section{Introduction}
Mesh routers and client devices are self-organized and self-configured to form wireless mesh networks(WMNs)
\cite{meshsurvey}.
A device is called a node in WMNs. Each node is equipped with multiple radios to improve the whole capacities in WMNs
\cite{yang0}.
The radios in WMNs are cognitive radios, by which the radio devices  are
capable of learning from their environment and adapting to the environment\cite{Cognitive}. Cognitive
radio is also called programmable radio because such radio has the ability of self-programming\cite{self},
 learning and reasoning \cite{Cognitive}.

Machine learning has been  studied for  about 60 years. It evolved from
simple  artificial intelligence to a wide variety of applications in image processing, vision, networking,
and pattern recognition.
%
In this paper, we propose a
learning algorithm for a forwarding node to find one of its links  with possibly maximal bandwidth, and then choose  next forwarding  node and then forward the message to that node. Each node only saves the last three changed bandwidth status of its links . Then the forwarding node learns the three status and predict the potential bandwidth of its links. So the forwarding node is able to
    find the neighbor with highest link bandwidth as its next hop.
We further devise an  algorithm to let the source node
figure out the forwarding region in order to avoid  flooding.

The rest of the paper is organized as follows. Section
II discusses the related research on this topic. Section III
proposes our novel forwarding method that selects the best next hop.
We evaluate the proposed schemes via simulations and
describe the performance results in Section IV. Section V
concludes the paper.

\section{Related Work}
Some approaches on machine learning, wireless forwarding and related work  have been studied \cite{yang1}\cite{MTA}\cite{yang2}\cite{yang3}\cite{shen1}\cite{shen2}\cite{shen3}.
Wang Y. et al. \cite{wang1}proposed a  machine learning mechanism  to
improve data transmission in sensor network. The predication of link quality was used
to implement the approach. Additionally, they developed  a protocol called MetricMap to
maintain efficient routing in case the regular routing is not working.


 Sawhney A. et al. \cite{Sawhney1} presented a machine learning algorithm to handle congestion  controlling  in  wireless  networks. Their approach
 learns many factors that have impact to congestion controlling, and then uses the parameters in a fuzzy logic to generate better result when congestion takes place. The efficiency is assessed with machine learning tools.

\section{The Learning Based Forwarding Mechanism}
\subsection{The Forwarding Problem}
In wireless mesh networks(WMNs), the communications  are over links. Link bandwidth is critical for transmission speed.
Since each node may be equipped with multiple network interfaces with different radios and the radios are switchable, the bandwidth over two neighbor nodes
may vary from time to time. The radios in WMNs are cognitive radios and then the nodes  are able to learn the changes of past bandwidths
and can further predict and select the desired link with potential highest bandwidth.

\begin{figure}[!htp]
\begin{center}
\includegraphics[width=6.0cm]{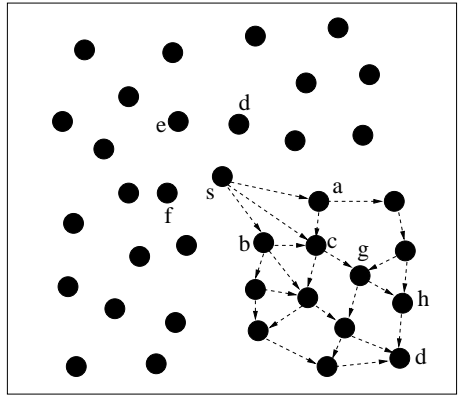}
\end{center}
\caption{Topology of a Wireless Mesh Network} \label{fig11}
\end{figure}

Assuming a source  node  $s$ intends to send data to a destination node $d$, many traditional routing algorithms set up the forwarding path
by simply selecting the shortest route. For example, $s-c-g-h-d$ is the forwarding path in figure 1. However, it may not be the best path in WMNs.
In WMNs, the bandwidth over two nodes changes frequently. The bandwidth of the link $sc$ is possibly much lower than that of $sa$. Or the past bandwidth of
$sc$ is higher than $sa$ but two much traffic is over $sc$ now so the available bandwidth of $sc$ is going down while that of $sa$ is going up.

Our goal is to let each  forwarding node select the link for next hop with
the highest potential bandwidth. In our approach, each node
learns its links' past bandwidths and then predict their potential bandwidths.
Then the forwarding node
figures out  its next hop with highest potential bandwidth.

\subsection{Prediction for Future Bandwidth}

Suppose node $i$ saves the bandwidth changes of its links of the last three times $t_{0}$, $t_{1}$, and $t_{2}$. Then for any of its neighbor  $j$, $i$ predicts the potential bandwidth of link $ij$. By computational method\cite{comp}, we define
 \begin{displaymath}
\alpha_{i,j,k}=
\sum_{m=0}^{k} {\frac{B_{i,j,m}}{\prod_{n=0}^{k}(t_{m}-t_{n})}} ~~~~  (1)
\end{displaymath}
 at time $t_{k}$, where $B_{i,j,m}$ is the bandwidth between node $i$ and node $j$ at time $m$. Then the bandwidth of link $ij$ at future time $p$ can be calculated and predicated as:\\

  \begin{displaymath}
 B_{i,j,p}=\alpha_{i,j,0}+\alpha_{i,j,1}(t_{p}-t_{0})+\alpha_{i,j,2}(t_{p}-t_{1})(t_{p}-t_{0}) ~~~~   (2)
 \end{displaymath}

Algorithm 1 describes  node $i$  learns the bandwidth of link $ij$ in the last three changes and then it predicts
the bandwidth of next time $p$.

  \begin{algorithm}[]
\begin{algorithmic}[1]
\caption{{\em Prediction for future bandwidth of link $ij$}} \label{algorithm4}
\STATE Learn and keep the bandwidth changes of link $ij$ at the last three times 0,1, and 2.
\STATE Calculate $\alpha_{i,j,k}$ with equation (1)
\STATE Calculate the predicted bandwidth of link $ij$ of time $p$ with equation (2)
\end{algorithmic}
\end{algorithm}

\subsection{Forwarding Region}
When a node $s$ intends to send data to
node $d$, it selects the neighbor node with highest potential
link bandwidth as its next hop and then same metric continues to select the best next forwarding node . Apparently, $s$ will not select any
nodes in the opposite direction from $s$ to $d$.
How is node $s$ aware of  the region where the next hop
falls? In current WMNs, each device is equipped with GPS
and hence it knows its location. We assume that the sender
knows its own location and the location of the receiver. The
assumption is very common in geographic routing\cite{yang1}.
Figure 2 shows the scenario. Suppose node $s$ intends to send
 data to node $d$, it figures out the forwarding region as algorithm 2.

 \begin{algorithm}[]
\begin{algorithmic}[1]
\caption{{\em Figure out the region for next hop}} \label{algorithm2}
\STATE $s$ connects $d$.
\STATE $sd$ rotates 45 degrees anti-clockwise, the ray is the positive half of $X$ axis.
\STATE $sd$ rotates 45 degrees clockwise, the ray is the negative half of $Y$ axis.
\STATE Oppositely extends the ray of $X$ axis to generate the negative half of $X$ axis.
\STATE Oppositely extends the ray of $Y$ axis to generate the positive half of $Y$ axis.
\STATE The plane is divided up to  four quadrants. The 4th quadrant is where the forwarding will be conducted.

\end{algorithmic}
\end{algorithm}

\begin{figure}[!htp]
\begin{center}
\includegraphics[width=6.5cm]{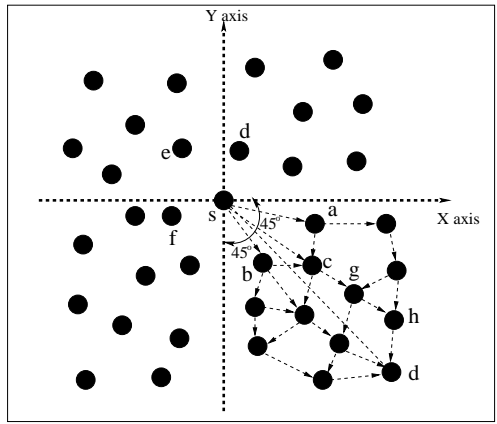}
\end{center}
\caption{Forwarding Region} \label{fig22}
\end{figure}


\subsection{Forwarding algorithm}
Suppose each node in a Wireless Mesh Network regularly mains the last three changes of bandwidths of all its links that connect its neighbors.
When node $s$ intends to send data to node $d$, $s$ first uses algorithm 2 to figure out the region where the forwarding
will be performed. Then $s$ calls algorithm 1 to find the node with potential highest bandwidth among all its neighbors as next hope. When the selected node relays
the forwarding, it only considers its neighbors in the forwarding  region as forwarding candidates, and it calls algorithm 1 to forward the data to next hop with potential
highest  bandwidth. The forwarding resumes until the packets arrive destination node $d$.

\begin{algorithm}[]
\begin{algorithmic}[1]
\caption{{\em Forwarding algorithm}} \label{algorithm2}
\STATE $s$ calls algorithm 2 to figure out the forwarding region.
\STATE $s$ calls algorithm 1 to find the node $n$ with potential highest bandwidth of link $sn$ as next hope, where $n$ is in the forwarding region.
\STATE if $n$ is $d$, end the algorithm. Otherwise $s$=$n$, go to step 2.

\end{algorithmic}
\end{algorithm}
\section{Evaluation}
We evaluated our mechanism in a simulated noiseless radio
network environment by MATLAB. We create a topology
that consists of a number of randomly distributed nodes.  We compare our approach("ML Forwarding") with two other algorithms.
One is congestion control and fuzzy logic with machine learning for wireless communications, say Fuzzy Logic. The other one is
supervised learning approach for routing optimization in wireless networks, say Supervised Learning.
The compared metrics are transmission delay(Milliseconds) and transmission speed(MBs/Millisecond).
We performed a sequence of experiments in which the number of
nodes varies from 100 to 300 in increments of 25 over an area of 100x100 meters in the reference network. For each
number of mobile users, we conduct our experiments 10 times and
present the average value.

\begin{figure}[!htp]
\begin{center}
\includegraphics[width=7.5cm]{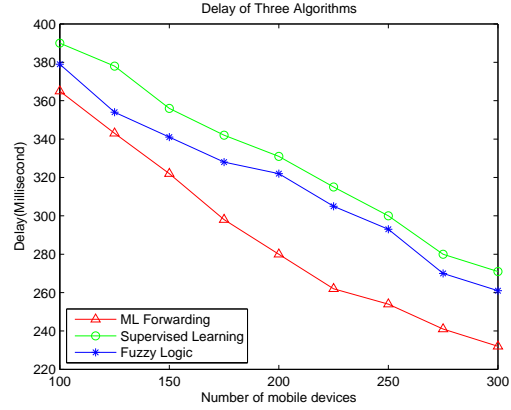}
\end{center}
\caption{Transmission Delay of the Three Algorithms} \label{fig13}
\end{figure}

Figure 3 shows that our approach results in the least delay. It is because our approach selects the link with potential
maximum bandwidth of each hop. Figure 4 shows that with the same reason, our approach generates the maximal transmission speed among the three approaches.

\begin{figure}[!htp]
\begin{center}
\includegraphics[width=7.5cm]{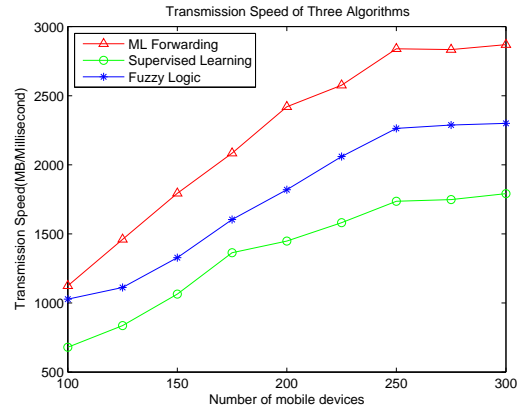}
\end{center}
\caption{Transmission Speed of the Three Algorithms}
\label{fig14}
\end{figure}

\section{Conclusion}
A machine learning based forwarding algorithm in wireless mesh networks with cognitive radios  is presented in this paper.
In this algorithm, each mobile device keeps the last three times of bandwidth changes of its links that connect its neighbors. Then when a node
 intends to forward data, the node learns the historical changes of bandwidth and then predicts the possible future bandwidths
  of the links with neighbor nodes. Hence the forwarding node is able to select the next hop with highest bandwidth.
  We also designed a geometrical algorithm to let the source node figure out the forwarding region in order to avoid unnecessary flooding.
 Simulation results demonstrate that our approach
outperforms peer approaches. \\
~~~\\
~~~\\
\large{ACKNOWLEDGMENT}\\

\normalsize
{This work  is supported in part by the Spanish government, Direcci\'{o}n General de Investigación Científica y T\'{e}cnica, a unit of the Ministerio de Econom\'{i}a y Competitividad, TIN2015-69542-C2-1-R (MINECO/FEDER), in collaboration with Universidad Rey Juan Carlos, Spain, under the project ``Inteligencia Artificial y M\'{e}todos Matem\'{a}ticos Avanzados para el Reconocimiento Autom\'{a}tico de Actividades.''}

%

%
%
%
%
%
%
%

%
\end{document}